\documentclass[%
reprint,
superscriptaddress,
amsmath,amssymb,
aps,
prl,
floatfix,
longbibliography
]{revtex4-2}

\usepackage{times}
\usepackage{amstext}
\usepackage{graphicx}
\usepackage{color}
\usepackage{soul}
\usepackage{MnSymbol}
\usepackage{subfiles}
\usepackage{bm}
\usepackage{hyperref}
\usepackage{layouts}
\usepackage{enumitem}
\usepackage{natbib}
\usepackage{sgame}
\usepackage{comment}
\usepackage{xcolor}

\begin{document}

\title{Group-Level Imitation May Stabilize Cooperation}

\author{Pierre Bousseyroux}
\affiliation{Chair of Econophysics and Complex Systems, École Polytechnique, 91128 Palaiseau Cedex, France}
\affiliation{LadHyX UMR CNRS 7646, École Polytechnique, Institut Polytechnique de Paris,  91128 Palaiseau Cedex, France}

\author{Gilles Zérah}
\affiliation{Capital Fund Management, 23 Rue de l’Université, 75007 Paris, France}

\author{Michael Benzaquen}
\affiliation{Chair of Econophysics and Complex Systems, École Polytechnique, 91128 Palaiseau Cedex, France}
\affiliation{LadHyX UMR CNRS 7646, École Polytechnique, Institut Polytechnique de Paris,  91128 Palaiseau Cedex, France}\affiliation{Capital Fund Management, 23 Rue de l’Université, 75007 Paris, France}

\begin{abstract}
Stabilizing cooperation among self-interested individuals presents a fundamental challenge in evolutionary theory and social science. While classical models predict the dominance of defection in social dilemmas, empirical and theoretical studies have identified various mechanisms that promote cooperation, including kin selection, reciprocity, and spatial structure. In this work, we investigate the role of localized imitation in the evolutionary dynamics of cooperation within an optional Public Goods Game (PGG). We introduce a model where individuals belong to distinct groups and adapt their strategies based solely on comparisons within their own group. We identify different dynamical regimes, including stable fixed points, limit cycles, and Rock-Scissors-Paper-type oscillations. Our analysis, grounded in a replicator-type framework, reveals that such group-level imitation can stabilize cooperative behavior, provided that groups are not initially polarized around a single strategy. In other words, restricting imitation to group-level interactions mitigates the destabilizing effects of global competition, providing a potential explanation for the resilience of cooperation in structured populations. 
\end{abstract}

\date{\today}

\maketitle

The emergence and persistence of cooperation among self-interested individuals have long posed a central problem in evolutionary theory and social science. As early as the 1980s, Axelrod and Hamilton emphasized the paradoxical nature of cooperation in a Darwinian framework, where short-term incentives favor defection \cite{axelrod1981evolution}. Several mechanisms have since been proposed to explain altruistic behavior, including kin selection \cite{wd1963evolution}, group selection (often formalized as multi-level selection) \cite{wilson1994reintroducing}, and reciprocal altruism \cite{trivers1971evolution}.

Evolutionary game theory, extended by Axelrod, Maynard Smith, and others \cite{axelrod1981evolution,nowak2006five}, provides a powerful framework for understanding how strategies spread through imitation, learning, or selection. The replicator equation \cite{hofbauer1998evolutionary} and stochastic processes such as Moran or Wright--Fisher \cite{chalub2006continuous,nowak2004emergence,taylor2004evolutionary,imhof2006evolutionary} offer insights into these dynamics, even when agents rely on local rather than global information.

A classical paradigm for studying cooperation is the Public Goods Game (PGG), a multi‐player extension of the Prisoner’s Dilemma \cite{fehr2002altruistic,hauert1997effects}. In PGGs, individuals choose whether to contribute to a public good that is then multiplied and shared among all participants. Under standard rationality, defection dominates—leading to the “tragedy of the commons” \cite{hardin2018tragedy}—yet empirical evidence indicates that real individuals contribute more than predicted \cite{dugatkin1997cooperation}. Mechanisms such as punishment have been proposed to account for this discrepancy \cite{fehr2000cooperation,brandt2005punishing}.

Another strategy to foster cooperation is to allow voluntary participation. The optional Public Goods Game (OPGG) proposed by Hauert and colleagues \cite{hauert2002volunteering,hauert2002replicator} adds the possibility of opting out, thereby introducing loners alongside cooperators and defectors. Although the loner strategy often prevents cooperators from being fully exploited, it typically does not yield a stable high fraction of cooperators. Numerous refinements have thus been explored, including minimal penalties \cite{sasaki2012take}, reputation systems \cite{righi2021reputation,capraro2021reputation}, and more sophisticated behavioral strategies \cite{podder2023complexity}. Other studies have highlighted the role of punishment timing \cite{rand2011evolution,quan2019benefits} or ostracism \cite{nakamaru2014effect}.

Beyond individual-level mechanisms, population structure plays a crucial role. Network models demonstrate that cooperators tend to cluster, which limits exploitation by defectors \cite{castellano2009statistical,szabo2007evolutionary}, while multi-population or interdependent network settings \cite{wang2012probabilistic,cooney2023evolution,hu2023complex} support cooperation through self-organized cycles. In these models, agents rely on local interactions rather than global comparisons.

In the present work, we investigate how \emph{group-level imitation} influences the evolution of cooperation in an optional Public Goods Game. Contrary to classical models where payoffs are compared across the entire population, we assume that imitation occurs \emph{exclusively within each subpopulation}, with no cross-group comparisons. This \emph{multi-population} approach resonates with recent studies on structured populations \cite{cooney2023evolution,hu2023complex,liu2024multi}, where group membership defines the domain of interaction. This corresponds to the case where two populations with identical characteristics share the same space. By integrating \emph{replicator-type dynamics}, we show that cooperation can be stabilized \emph{as long as the groups are not initially polarized} (i.e., dominated by a single strategy). Our results thus reveal how the interplay between \emph{voluntary participation} and \emph{group-level imitation} can foster evolutionary stability in structured populations.

\paragraph*{Model.}

Our model's starting point is given by the work of ~\citet{hauert2002replicator}. Consider two large populations of individuals, of sizes $N_1$ and $N_2$, that never mix. 
Each individual in these groups is assigned a fixed strategy among the three following options: cooperators, defectors, and loners.

We denote by $x_i, y_i,$ and $z_i$ the proportions of cooperators, defectors, and loners, respectively, in group $i \in \{1, 2\}$. One naturally has:
\begin{equation}\label{normalisation}
    x_i + y_i + z_i = 1
\end{equation}
for $i \in \{1, 2\}$.
From time to time, a group of $m$ players is randomly sampled from the entire population to play an optional PGG. This group may contain individuals from both groups. A common pool is established, to which only cooperators contribute, while both cooperators and defectors benefit from it. Each cooperator contributes a fixed amount $c$ to the common pool. If there are $m_c$ cooperators in the group, the total contribution to the pool is thus $c\,m_c$. This amount is then multiplied by a factor $r$ and evenly redistributed among all active participants (cooperators and defectors). In contrast, loners do not participate in the game and automatically receive a fixed payoff  $\sigma c$. At the end of a single round, the net payoffs are given~by:
\begin{equation}
    P^x = -c + rc\frac{m_c}{m}, 
    \quad
    P^y = rc\frac{m_c}{m}, 
    \quad
    P^z = \sigma c,
\end{equation}
where $P^x$, $P^y$, and $P^z$ denote the net payoffs of cooperators, defectors, and loners, respectively, within the selected group of $m$ individuals.
Without loss of generality,  we henceforth set $c=1$ by selecting a suitable unit of currency.

In practice, such optional PGGs---each involving a randomly selected set of $m$ players---are played repeatedly. We denote by $\tau_g$ the time interval between successive games. In addition to these games, an adaptation mechanism takes place: at time intervals of length $\tau_a$, two individuals, denoted $j$ and $k$, are randomly selected within the same group. Their most recent payoffs, $P^j$ and $P^k$, are then compared. In that case, individual $k$ adopts the strategy of $j$ with probability $\beta (P^j - P^k)_+$ (and remains unchanged with probability $1 - \beta (P^j - P^k)_+$), where $\beta$ is a fixed proportionality constant, while individual $j$ retains their own strategy. Let us stress again that no cross-group comparisons occur.

If we now consider that groups form randomly, we obtain for the continuous time evolution of the strategies in each group, $i \in \{1,2\}$, the replicator dynamics (see Section A in SI):
\begin{equation}\label{1e}
\begin{aligned}
    \dot{x}_i &= \frac{x_i}{\tau}\,\Bigl(\overline{P}^x - \overline{P}_i\Bigr),\\[1mm]
    \dot{y}_i &= \frac{y_i}{\tau}\,\Bigl(\overline{P}^y - \overline{P}_i\Bigr),\\[1mm]
    \dot{z}_i &= \frac{z_i}{\tau}\,\Bigl(\overline{P}^z - \overline{P}_i\Bigr),
\end{aligned}
\end{equation}
where the average payoff in group $i$ is given by
\begin{equation}
    \overline{P}_i = x_i\,\overline{P}^x + y_i\,\overline{P}^y + z_i\,\overline{P}^z.
\end{equation}
We note that the average payoff of each strategy $\overline{P}^x$, $\overline{P}^y$, and $\overline{P}^z$ depend only on the fractions $x$, $y$, and $z$ of the three strategies 
and may be calculated analytically (see Section B in SI):
\begin{equation}\label{2e}
\begin{aligned}
    \overline{P}^y &= \sigma z^{m-1} + \frac{rx}{1 - z} \left(1 - \frac{1 - z^m}{m(1 - z)}\right),\\[1mm]
    \overline{P}^x &= \overline{P}^y - f(z),\\[1mm]
    \overline{P}^z &= \sigma.
\end{aligned}
\end{equation}
Here, $x$ and $z$ denote the fractions of cooperators and loners in the global population, respectively, and are defined as
\begin{equation}\label{defxz}
    x = \alpha x_1 + (1-\alpha) x_2 \quad \text{and} \quad z = \alpha z_1 + (1-\alpha) z_2,
\end{equation}
with $\alpha = \frac{N_1}{N_1 + N_2}$, and
\begin{equation}
    f(z) = 1 + z^{m-1}(r-1) - \frac{r(1 - z^m)}{m(1 - z)}.
\end{equation}

In the replicator equation \eqref{2e}, the payoff of each strategy is compared to the average payoff of the strategies \emph{within each group}. The payoff part is global to the system, while the selection part occurs within the group: this corresponds to the situation where two groups with identical characteristics share the same environment.

\paragraph*{Numerical Results.}

\begin{figure}[t!]
    \centering
    \includegraphics[scale = 1]{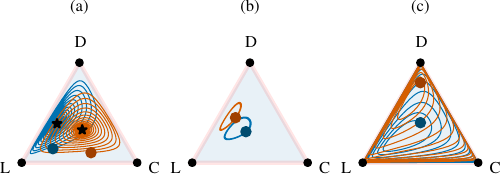}
    \caption{Numerical trajectories $(x_i,y_i,z_i)$ of the system~\eqref{1e}, visualized in the simplex $x+y+z=1$, are plotted in blue for $i=1$ and in orange for $i=2$. The three vertices correspond to pure strategies: Cooperators $C$ at $(1,0,0)$, Defectors $D$ at $(0,1,0)$, and Loners $L$ at $(0,0,1)$. 
    We use parameters $m=7$, $\sigma=1$, $r=4$, $\tau=1$, and $\alpha=0.7$. 
    Each panel (a), (b), and (c) shows two trajectories (one for each of two groups) starting from distinct initial conditions, indicated by dots: (a)~$\,(x_1, y_1, z_1) = (0.2, 0.14, 0.66)$ and $(x_2, y_2, z_2) = (0.55, 0.1, 0.35)$,     (b)~$\,(x_1, y_1, z_1) = (0.31, 0.31, 0.38)$ and $(x_2, y_2, z_2) = (0.15, 0.45, 0.4)$,
    (c)~$\,(x_1, y_1, z_1) = (0.3, 0.4, 0.3)$ and $(x_2, y_2, z_2) = (0.1, 0.8, 0.1)$. In panel (a), the two fixed points toward which the trajectories converge are represented by black stars.
    }
    \label{fig:DP1}
\end{figure}

\noindent
We numerically integrate the system~\eqref{1e} (using Eqs.~\eqref{2e}) by means of a fourth-order Runge--Kutta (RK4) scheme \cite{kutta1901beitrag}. Three distinct sets of initial conditions are considered, each giving rise to a different dynamical outcome (see Figure~\ref{fig:DP1}): 
(a)~convergence to a fixed point (with trajectories spiraling around it), (b)~closed periodic orbits and (c)~an heteroclinic limit cycle of the rock--paper--scissors-type at the boundary of the simplex.

\paragraph{Analytical Results. } 
We are interested in the fixed points inside the simplex $x + y + z = 1$. According to Eqs.~\eqref{1e}, the quantities 
$\overline{P}_i^x, \overline{P}_i^y, \overline{P}_i^z$
must converge to the same value, which can only be $\sigma$. 
Setting $\overline{P}^x_i=\overline{P}^y_i=\sigma$ in Eq.~\eqref{2e} yields:
\begin{equation}\label{fix}
\begin{aligned}
\sigma &= \sigma\, z^{\,m-1} + \dfrac{r\, x}{1 - z}\,\left(1 - \dfrac{1 - z^m}{m(1-z)}\right)\!,\\
0 &= f(z).
\end{aligned}
\end{equation}
Denoting $x^*$ and $z^*$ the solutions of this system, all the individual fractions 
$x_1, y_1, z_1, x_2, y_2, z_2$
that satisfy
\begin{equation}
    \alpha x_1 + (1-\alpha)x_2 = x^*
    \quad \text{and} \quad
    \alpha z_1 + (1-\alpha)z_2 = z^*,
\end{equation}
as well as the normalization conditions \eqref{normalisation}, thus correspond to fixed points inside the simplex. 

From Eqs.~\eqref{1e}, one can show that 
\begin{equation}\label{rapports}
    q_{x, y} := \frac{x_2 y_1}{y_2 x_1},
    \quad
    q_{x, z} := \frac{x_2 z_1}{z_2 x_1}
\end{equation}
are constants of motion, thereby providing two independent conserved quantities.
Thus, once $m$, $\sigma$, and $r$ are fixed, the values $x^*$ and $z^*$ are consequently determined, defining a two-dimensional manifold of possible  fixed points. To specify which particular fixed point the system may eventually reach, it is sufficient to know the initial values of $q_{x, y}$ and $q_{x, z}$.

Now that we have identified the fixed points of the system, denoted $(x_1^*, y_1^*, z_1^*, x_2^*, y_2^*, z_2^*)$ we analyze their stability by linearizing the dynamics around them:
\begin{equation}\label{Adef}
\begin{pmatrix}
\dot{\delta x}_1 \\
\dot{\delta y}_1 \\
\dot{\delta z}_1 \\
\dot{\delta x}_2 \\
\dot{\delta y}_2 \\
\dot{\delta z}_2
\end{pmatrix}
=
A
\begin{pmatrix}
\delta x_1 \\
\delta y_1 \\
\delta z_1 \\
\delta x_2 \\
\delta y_2 \\
\delta z_2
\end{pmatrix}
\end{equation}where $A$ is the Jacobian matrix and
\begin{equation}
\delta x_i = x_i - x_i^*, \quad 
\delta y_i = y_i - y_i^*, \quad 
\delta z_i = z_i - z_i^*
\end{equation}for $i\in \{1, 2\}$.

\begin{figure}[t!]
    \centering
    \includegraphics[scale = 1]{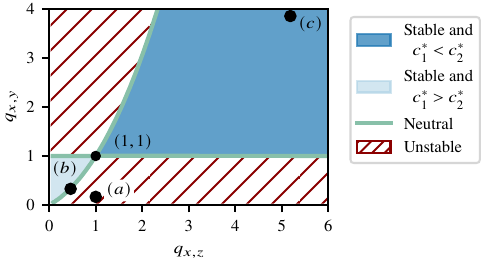}
    \caption{We analyze the sign of the common real part of the two complex conjugate eigenvalues of the matrix $A$ (defined in Eq.~\eqref{Adef}) as a function of the ratios $q_{x,y}$ and $q_{x,z}$ (defined in Eq.~\eqref{rapports}), with parameters $m=7$, $\sigma=1$, $r=4$, $\tau=1$, and $\alpha=0.7$. In the figure, regions where the real part is positive are depicted with a hatched pattern, while the green boundary highlights the locus where the real part vanishes, i.e., where the fixed point is neutral. The region where the real part is negative—which corresponds to convergence towards a fixed point—is further subdivided into two connected components according to the Boolean condition $c_1^* < c_2^*$, with the quantities $c_1^*$ and $c_2^*$ defined in Eq.~\eqref{c}. We also indicate the positions of the three values of the pair $(q_{x,z}, q_{x,y})$ used in Fig.~\ref{fig:DP1}, corresponding to the desired regimes.}
    \label{fig:newfigure}
\end{figure}

\begin{figure}[t!]
    \centering
    \includegraphics[scale = 1]{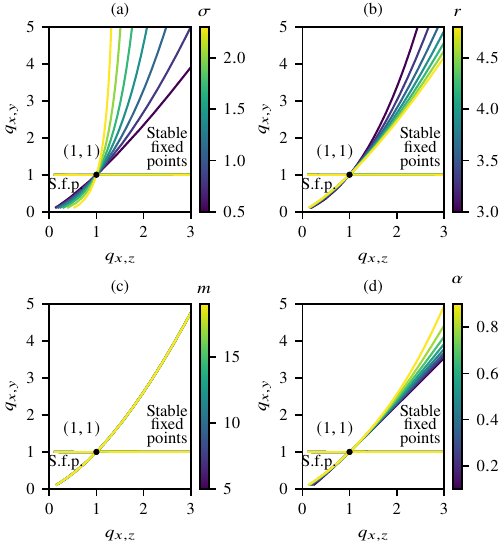}
    \caption{We investigate the sign of the common real part of the two complex conjugate eigenvalues of the matrix $A$ (defined in Eq.~\eqref{Adef}) as a function of the ratios $q_{x,y}$ and $q_{x,z}$ (defined in Eq.~\eqref{rapports}) for different parameters $(r, m, \alpha, \sigma)$. Numerically, the region where the real part is negative is indicated in the plots as “Stable fixed points” (abbreviated as S.f.p.). The baseline parameters are $r=4$, $m=7$, $\alpha=0.7$, and $\sigma=0.75$. In panels (a)–(d), the effect of varying one parameter—(a) $\sigma$, (b) $r$, (c) $m$, and (d) $\alpha$—on the two boundaries of the stable fixed points region is shown, with the other parameters fixed at their baseline values.}
    \label{fig:frontiere30}
\end{figure}

Due to the normalization Eqs.~\eqref{normalisation} and the existence of two constants of motion defined in Eqs.~\eqref{rapports}, $A$ is a rank-$2$ matrix. Generically, it has two nonzero, complex-conjugate eigenvalues. Looking at the real part of these eigenvalues as a function of the ratios $q_{x, y}$ and $q_{x, z}$ fully characterizes the stability of the fixed point. Such analysis reveals two regimes:
\begin{itemize}
    \item A \emph{stable} regime, where the system converges to a fixed point and the real part of the eigenvalues is negative. See Fig.~\ref{fig:DP1}(a). 
    \item An \emph{unstable} regime, where the real part is positive and the system exhibits a heteroclinic Rock-Paper-Scissors-type limit cycle. See Fig.~\ref{fig:DP1}(c).    
\end{itemize}
These two regions are separated by a one-dimensional boundary along which the system exhibits two stable periodic orbits (see Fig.~\ref{fig:DP1}(b)). This boundary appears to consist of two branches: one corresponding to $q_{x,y} = 1$, and another less trivial branch whose shape depends on the parameters $\sigma$, $r$, and $\alpha$. (see Fig.~\ref{fig:frontiere30}). The region from which trajectories converge to a fixed point is always divided into two distinct domains. An interesting quantity 
 to distinguish between them numerically is to introduce the \textit{effective net added value per capita} of group $i$ at the fixed point, defined as  
\begin{equation}\label{c}
    c_i^* = (r - 1)x_i^* - \sigma z_i^*,
\end{equation} for $i \in \{1, 2\}$~\footnote{Note that the \textit{effective net added value per capita} of the entire population $c := (r - 1)x - \sigma z$ within convergence zones reaches a stationary value $c^* := (r - 1)x^* - \sigma z^*$ independently of initial conditions.}. 
As shown in Fig.~\ref{fig:newfigure}, based on the same parameters as in Fig.~\ref{fig:DP1}, the conditions $c_1^* > c_2^*$ and $c_2^* > c_1^*$ effectively distinguish the two convergence zones. Numerical evidence indicates that this distinction—$c_1^* > c_2^*$ for the lower-left zone and $c_2^* > c_1^*$ for the upper-right—holds consistently across the entire parameter range explored in Fig.~\ref{fig:frontiere30}.

\paragraph{Discussion.} 
The case $ \alpha = 1 $, meaning that the population consists entirely of group 1, has been fully analyzed by~\citet{hauert2002replicator}. It exhibits a Rock-Scissors-Paper type cyclic dominance: a fixed point is surrounded by closed orbits such that the system exhibits stable periodic oscillations of the strategy frequencies. Such cycles can be understood as follows. If cooperators are in the majority, the temptation to defect increases because defectors benefit from a well-funded pool without paying the cost of cooperation. However, when defection becomes widespread, the pool depletes, making the solitary strategy more attractive. A reduction in the number of participants decreases the dilution of gains among the remaining cooperators, making cooperation viable again. Thus, there is never convergence to a fixed point. The question we have addressed here is: What happens if, instead of comparisons being made within the entire group, they occur between smaller groups?

First, our observations reveal a novel regime characterized by convergence. As illustrated in panel~(a) of Fig.~\ref{fig:DP1}, the proportions of strategists in each group stabilize over time. This behavior - absent from the original results of ~\citet{hauert2002replicator} - depends on the initial values of $q_{x,y}$ and $q_{x,z}$, as highlighted by the stable fixed-point regions in Fig.~\ref{fig:frontiere30}. Numerical evidence suggests that this regime emerges when $ 1 < q_{x,y} \lesssim q_{x,z}  $ or $  q_{x,z} \lesssim q_{x,y} < 1 $  (see Fig.~\ref{fig:newfigure}). Note that if either $q_{x,y}$ or $q_{x,z}$ approaches zero (but not both simultaneously), these inequalities cannot be satisfied. According to their definitions in Eq.~\eqref{rapports}, this intuitively corresponds to a scenario where one group is initially polarized toward a single strategy. Since these ratios remain constant over time, polarization persists indefinitely, preventing convergence toward cooperative equilibria.

Furthermore, \citet{hauert2002replicator} previously showed that increasing $\sigma$ or decreasing $r$ promotes cooperation. Our findings further demonstrate that this further enhances convergence towards stable fixed points by expanding convergence regions (Fig.~\ref{fig:frontiere30}, panels~(a) and~(b)). Interestingly, the convergence regions are insensitive to changes in group size $m$, provided that it is larger than a given $m_{\min}$ that depends on $r$  (see Fig.~\ref{fig:frontiere30}(c)), indicating that the number of participants in the optional public goods game has no impact on the stabilization of cooperation. Moreover, panel~(d) of Fig.~\ref{fig:frontiere30} reveals that as $\alpha$ approaches $1$ (resp. 0.5), unstable regions shrink (resp. widen), signifying that cooperation becomes more robust against initial polarization when one group clearly dominates.

Moreover, the cyclic dynamics regime appears along two distinct one-dimensional branches depicted in Fig.~\ref{fig:frontiere30}, intersecting at the symmetric point $(1,1)$. Such an intersection is expected, as it corresponds precisely to the situation in which both groups satisfy identical inequality conditions, thus reducing the dynamics to the classical system studied by ~\citet{hauert2002replicator}. Due to the inherent symmetry between groups 1 and 2, our analysis must remain invariant under the transformation $(q_{x,y}, q_{x,z}, \alpha) \mapsto (1/q_{x,y},1/q_{x,z},1-\alpha)$, as implied by Eq.~\eqref{rapports}. Numerically, this symmetry is indeed observed, notably in Fig.~\ref{fig:frontiere30}. In the case where $\alpha = 0.5$, the boundary curves must follow power laws of the form $q_{x, y} = q_{x, z}^\beta$, as imposed by symmetry: $\beta = 1$ for the branch where $q_{x,y} = 1$, and $\beta \approx 1.5$ for the other. It is interesting that these idealized boundaries, representing infinitely rational agents, remain easily observable in numerical simulations (see panel~(b) of Fig.~\ref{fig:DP1}). Further research is necessary to clarify how characteristic times scale as the system approaches these cyclic boundaries.

In conclusion, our results demonstrate that the introduction of group-level imitation mechanisms helps stabilize cooperation, provided that a group is not initially too polarized toward a single strategy. In such cases, although these mechanisms do not quantitatively enhance cooperation — since the long-term proportions of loners and cooperators in the total population are independent of initial conditions— they nonetheless contribute to the convergence of the strategy proportions within each group. Moreover, our analysis naturally generalizes to systems composed of $k>2$ groups (see the preliminary discussion in Section~C of the SI), and further research is needed to fully characterize the dynamics of this generalized setting.

\paragraph*{Acknowledgements.} 

We are grateful to Thomas Valade for his active involvement in discussions and for his valuable insights. This research was conducted within the Econophysics
\& Complex Systems Research Chair, under the aegis of
the Fondation du Risque, the Fondation de l’Ecole polytechnique, the Ecole polytechnique and Capital Fund
Management.

\bibliographystyle{apsrev4-2} 
\bibliography{PGG}  

\clearpage
\onecolumngrid

\section*{Supplemental material}

\subsection{A: proof of equations~\eqref{1e}}

\noindent

The replicator equation may be established in many ways, e.g., from a flow equation (cf.~\citet{sandholm2008deterministic}), where the change in the proportion of any strategy is given by the difference between inflows and outflows per unit time, that is:
\begin{equation}
\dot x^j = \sum_{k=1}^n x^k \rho_{kj} - \sum_{k=1}^n x^j \rho_{jk},
\end{equation}
where $n$ denotes the number of strategies, and $\rho_{jk}$ is the switching rate from strategy $j$ to $k$. This quantity is the product of $x_j$ the probability to choose $j$ by the transition probability $(\overline{P}^j - \overline{P}^k)_+/\tau$ , where $ _+$ denotes the positive part. Since $x_+ - (-x)_+ = x$, we obtain:
\begin{equation}
\dot{x}^j = \sum_{k=1}^n x^j x^k \frac{(\overline{P}^j - \overline{P}^k)}{\tau} = \frac{x^j}{\tau} \left(\overline{P}^j - \overline{P}\right),
\end{equation}
where $\overline{P} = \sum_{k=1}^n x^k \overline{P}^k$. By considering this equation within a group $i$ and using our notation $(x^1, x^2, x^3) = (x, y, z)$ and $(\overline{P}^1, \overline{P}^2, \overline{P}^3) = (\overline{P}^x, \overline{P}^y, \overline{P}^z)$, we obtain equation~\eqref{1e}.

Alternatively, a fully microscopic derivation is possible (cf.~\citet{traulsen2005coevolutionary}) starting from a finite population description. The fraction of each strategy can be shown to obey a Langevin equation, which reduces to the deterministic mean-field equation~\eqref{1e} when the number of agents goes to infinity.

For a slightly different approach, we define the characteristic timescale over which the system's state evolves, denoted by $\tau$, as $\tau = N \tau_a \beta^{-1}$, where $N = N_1 + N_2$ is the total population size. We make two assumptions: (i) $\tau_a \gtrsim N \tau_g$, i.e., the imitation process is less frequent than the games, and (ii) $\tau_a \ll \tau$, which allows us to consider a time interval $\Delta t$ such that
\begin{equation}
\tau_a \ll \Delta t \ll \tau = \frac{N \tau_a}{\beta},
\end{equation}
which is possible since $\tau_a \ll \tau$. Consider $b \in \{x, y, z\}$ and $i \in \{1, 2\}$. Over this time interval $\Delta t$, there are on the order of $\frac{\Delta t}{\tau_a}$ comparisons—a number that is very large. Hence, by the law of large numbers, the variation in the number of individuals in group $i$ adopting strategy $b$ is given by
\begin{equation}\label{dis}
    \Delta n^b_i := \beta\,\frac{\Delta t}{\tau_a}\,b_i\Bigl(\overline{P}^b_i - \overline{P}_i\Bigr),
\end{equation}
where the average quantities are computed over the time period $\Delta t$. In order to write this expression, it is also necessary to assume that each individual has a recently updated payoff over the interval $\Delta t$ (i.e., individuals must have played). Therefore, we must have
\begin{equation}
    \Delta t \gg N\tau_g,
\end{equation}
which holds since $\tau_a \gtrsim N\tau_g$ and $\Delta t \gg \tau_a$. We also assume that the characteristic time scale for the evolution of the system's macroscopic quantities—such as $\overline{P}^b_i$, $\overline{P}_i$, and the strategy proportions—is much larger than $\Delta t$. We will verify this later. Let us now move on to a continuous approach by setting 
\begin{equation}
\tau := \frac{N\tau_a}{\beta}.
\end{equation}
Assuming that $\Delta t \ll \tau$, equation~\eqref{dis} becomes
\begin{equation}
    \dot{b_i} = \frac{b_i}{\tau}\,\Bigl(\overline{P}^b_i - \overline{P}_i\Bigr),
\end{equation}
with the mesoscopic time scale $\Delta t$ effectively disappearing. Thus, the characteristic time scale for the evolution of the system, from a macroscopic point of view, is given by $\tau$, which is much larger than $\Delta t$, the time window over which the averages are computed.

\subsection{B: proof of equations~\eqref{2e}}\label{proof2e}

Recall that the mean payoffs $\overline{P}^x_i$, $\overline{P}^y_i$, and $\overline{P}^z_i$ are computed over a time window $\Delta t$ satisfying $\tau_a \ll \Delta t \ll \tau$. Since $\Delta t \ll \tau$, these mean payoffs can be obtained by calculating the expected value of the payoff received by a randomly selected individual in a game, conditioned on the state of the population with fixed proportions $x_i$, $y_i$, and $z_i$.

Let us denote by $p_1$ the probability that an individual is an active player (i.e., a cooperator or defector) from group~1, by $p_2$ the probability that an individual is an active player from group~2, and by $z$ the probability that an individual is a loner (from either group). It is not difficult to see that
\begin{equation}
    \begin{cases}
        p_1 = \alpha\,(1 - z_1), \\
        p_2 = (1-\alpha)\,(1 - z_2), \\
        z = 1 - (p_1+p_2) = \alpha\,z_1 + (1-\alpha)\,z_2,
    \end{cases}
\end{equation}
where 
\begin{equation}
    \alpha = \frac{N_1}{N_1+N_2}
\end{equation}
is the probability of selecting an individual from group~1.

\vspace{1ex}

We now wish to calculate the average payoff of a random defector from group~1, denoted by $\overline{P}_1^y$. Let $S_1$ and $S_2$ be the random variables describing, respectively, the number of active players from groups~1 and~2 in a randomly selected group of $m$ individuals. Since a defector of group $1$ is already selected, the remaining $m-1$ individuals are chosen according to a multinomial law. In particular, the probability of obtaining $S_1-1$ players from group~1 and $S_2$ players from group~2, with the remaining $S_3 := m - (S_1+S_2)$ individuals being loners, is given by
\begin{equation}
    \mathbb{P}(S_1-1,\, S_2) = \frac{(m-1)!}{(S_1-1)! \; S_2! \; S_3!}\, p_1^{\,S_1-1}\, p_2^{\,S_2}\, z^{\,S_3},
\end{equation}
with the constraints $1\le S_1$, $0\le S_2$, and $S_1+S_2\le m$.

\vspace{1ex}

Let $m_1$ and $m_2$ denote the numbers of cooperators among the active players from groups~1 and~2, respectively. The corresponding conditional probabilities are
\begin{equation}
    \mathbb{P}(m_1 \mid S_1-1) = \binom{S_1-1}{m_1} \left(\frac{x_1}{x_1+y_1}\right)^{m_1} \left(\frac{y_1}{x_1+y_1}\right)^{S_1-1-m_1},
\end{equation}
and
\begin{equation}
    \mathbb{P}(m_2 \mid S_2) = \binom{S_2}{m_2} \left(\frac{x_2}{x_2+y_2}\right)^{m_2} \left(\frac{y_2}{x_2+y_2}\right)^{S_2-m_2}.
\end{equation}

\vspace{1ex}

In the optional public goods game, only active players (i.e., cooperators and defectors) contribute to and benefit from the common pool. However, if the selected group contains only one active player (i.e. $S_1+S_2 = 1$), then the defector of group $1$ is assumed to behave as a loner and receive a fixed payoff $\sigma$. Thus, we define the gain of the focal individual as
\begin{equation}
    G(m_1, m_2, S_1, S_2) :=
    \begin{cases}
        r\,\dfrac{m_1+m_2}{S_1+S_2}, & \text{if } S_1+S_2\ge2,\\[1mm]
        \sigma, & \text{if } S_1+S_2 = 1.
    \end{cases}
\end{equation}
The average payoff $\overline{P}_1^y$ for a cooperator in group~1 is then given by
\begin{equation}\label{eq:pi1y}
    \overline{P}_1^y = \sum_{\substack{S_1\ge 1,\, S_2\ge 0 \\ S_1+S_2\le m}}
    \sum_{\substack{m_1\le S_1-1,\, m_2\le S_2}}
    \mathbb{P}(S_1-1,\, S_2)\, \mathbb{P}(m_1 \mid S_1-1)\, \mathbb{P}(m_2 \mid S_2)\, G(m_1, m_2, S_1, S_2).
\end{equation}
One may decompose this expression as
\begin{equation}
    \overline{P}_1^y = \sigma\, z^{m-1} + A + B,
\end{equation}
where
\begin{equation}
    A = r\, \sum_{\substack{S_1\ge 1,\, S_2\ge 0 \\ 2\le S_1+S_2\le N}}
    \sum_{m_1\le S_1-1} \mathbb{P}(S_1-1,\, S_2)\, \mathbb{P}(m_1 \mid S_1-1)\,\frac{m_1}{S_1+S_2},
\end{equation}
and
\begin{equation}
    B = r\, \sum_{\substack{S_1\ge 1,\, S_2\ge 0 \\ 2\le S_1+S_2\le N}}
    \sum_{m_2\le S_2} \mathbb{P}(S_1-1,\, S_2)\, \mathbb{P}(m_2 \mid S_2)\,\frac{m_2}{S_1+S_2}.
\end{equation}
Since the sum over $m_1$ yields the expectation $\mathbb{E}(m_1\mid S_1-1) = (S_1-1)\frac{x_1}{x_1+y_1}$ (and similarly for $m_2$), we obtain
\begin{equation}
    A = \frac{r\,x_1}{x_1+y_1}\,\mathbb{E}\!\left(\frac{S_1-1}{S_1+S_2}\,\mathbf{1}_{\{S_1+S_2\ge2\}}\right),
\end{equation}
and
\begin{equation}
    B = \frac{r\,x_2}{x_2+y_2}\,\mathbb{E}\!\left(\frac{S_2}{S_1+S_2}\,\mathbf{1}_{\{S_1+S_2\ge2\}}\right).
\end{equation}
By further conditioning on the total number of active players, note that if we define
\begin{equation}
    X = S_1 + S_2 - 1,
\end{equation}
then $X$ follows a binomial distribution with parameters $m-1$ and $p_1+p_2$. One may show that
\begin{equation}
    A+B = \frac{r}{1-z}\Bigl(\alpha\,x_1 + (1-\alpha)\,x_2\Bigr)\left(1-\frac{1-z^m}{m(1-z)}\right),
\end{equation}
so that the average payoff of a cooperator in group~1 becomes
\begin{equation}\label{eq:pi1y_final}
    P_1^y = \sigma\,z^{m-1} + \frac{r}{1-z}\Bigl(\alpha\,x_1 + (1-\alpha)\,x_2\Bigr)\left(1-\frac{1-z^m}{m(1-z)}\right).
\end{equation}

\vspace{1ex}

Next, we compute the payoff difference between a cooperator and a defector in group~1, namely $\overline{P}_1^y-\overline{P}_1^x$. For this purpose, consider the gain function
\begin{equation}
    G(m_1, m_2, S_1, S_2) = \left(1 - \frac{r}{S_1+S_2}\right)\mathbf{1}_{\{S_1+S_2\ge2\}}.
\end{equation}
Defining again $X = S_1 + S_2 - 1$, where $X$ follows a binomial distribution with parameters $m-1$ and $p_1+p_2$, we obtain
\begin{equation}
    \overline{P}_1^y-\overline{P}_1^x = \mathbb{E}\!\left[\left(1 - \frac{r}{X+1}\right)\mathbf{1}_{\{X\ge1\}}\right]
    = \Bigl(1-z^{m-1}\Bigr) - r\,\mathbb{E}\!\left[\frac{1}{X+1}\mathbf{1}_{\{X\ge1\}}\right].
\end{equation}
Since
\begin{equation}
    \mathbb{E}\!\left[\frac{1}{X+1}\mathbf{1}_{\{X\ge1\}}\right]
    = \mathbb{E}\!\left[\frac{1}{X+1}\right] - z^{m-1}
    = \frac{1-z^m}{m(1-z)} - z^{m-1},
\end{equation}
we can substitute this back into the expression for $\overline{P}_1^y-\overline{P}_1^x$:
\begin{equation}
     \overline{P}_1^y-\overline{P}_1^x = \left(1-z^{m-1}\right) - r\left[\frac{1-z^m}{m(1-z)} - z^{m-1}\right].
\end{equation}
Expanding and regrouping the terms gives
\begin{align}
    \overline{P}_1^y-\overline{P}_1^x &= 1-z^{m-1} - \frac{r(1-z^m)}{m(1-z)} + r\,z^{m-1} \\
    &= 1 + z^{m-1}(r-1) - \frac{r(1-z^m)}{m(1-z)}.\label{diff}
\end{align}
This is the final expression for the payoff difference between a cooperator and a defector in group~1. Thus, equations~\eqref{eq:pi1y_final} and~\eqref{diff} lead to equations~\eqref{2e}.

\section{Generalization to $k$ groups}

Consider $k$ groups, each with $N_1,\dots,N_k$ individuals, and denote by $x_i$, $y_i$, $z_i$, $\overline{P}^x_i$, $\overline{P}^y_i$, and $\overline{P}^z_i$ (for $1\le i\le k$) the corresponding quantities. Equations \eqref{1e} and \eqref{2e} remain valid, with the global fractions $x$ and $z$ converging to $x^*$ and $z^*$, respectively. Since each group satisfies $x_i + y_i + z_i = 1,\quad i=1,\dots,k$, the system has $2k-2$ degrees of freedom. Moreover, the ratios
\begin{equation}
q_{x,y}^{i,j} := \frac{x_i y_j}{y_i x_j}\quad \text{and} \quad q_{x,z}^{i,j} := \frac{x_i z_j}{z_i x_j},\quad 1\le i\neq j\le k,
\end{equation}
are constants of motion. By selecting the independent set $q_{x,y}^{i,i+1}$ and $q_{x,z}^{i,i+1}$, we obtain $2(k-1)$ independent quantities—exactly matching the rank of the Jacobian. Thus, as in the two-group case, the final state is fully determined by these initial ratios together with the fixed points $x^*$ and $z^*$.

\end{document}